\documentclass[	pra,
				twocolumn,
				groupedaddress,
				showpacs,
				]{revtex4}

\usepackage{amsmath,amstext,amsfonts,amssymb,amsthm}
\usepackage{graphicx}				
\usepackage{bbold}
\usepackage{dsfont}					
\usepackage{enumerate}				
\usepackage{pgf}
\usepackage{tikz}
\usetikzlibrary{arrows,automata}
\def\ket#1{\left| #1\right>}
\def\bra#1{\left< #1\right|}
\def \beq {\begin{equation}}
\def \eeq {\end{equation}}
\newcommand{\vek}[1]{\mathbf #1}
\newcommand{\ketbra}[2]{|#1\rangle\!\langle #2|}

\newcommand{\ketbraIX}[3]{\ket{#1}_{\! #2}\!\!\bra{#3}}

\usepackage{ulem}
\usepackage[pdftex,bookmarks=true]{hyperref} 
				
\begin{document}
\title{Quantum superpositions of crystalline structures}
\author{Jens D. \surname{Baltrusch}$^{1,2}$}
\email[e-mail: ]{jens.baltrusch@physik.uni-saarland.de}
\author{Cecilia \surname{Cormick}$^{1}$}
\author{Gabriele \surname{De Chiara}$^{2,3}$}
\author{Tommaso \surname{Calarco}$^{4}$}
\author{Giovanna Morigi$^{1,2}$}
\affiliation{
$^1$ Theoretische Physik, Universit\"at des Saarlandes, D-66123 Saarbr\"ucken, Germany\\
$^2$ Grup d'Optica, Departament de Fisica, Universitat Aut\`onoma de Barcelona, E-08193 Bellaterra, Spain \\ 
$^3$ Centre for Theoretical Atomic, Molecular and Optical Physics, School of Mathematics and Physics, Queen's University, Belfast BT7 1NN, United Kingdom \\ 
$^4$ Institut f\"ur Quanteninformationsverarbeitung, Universit\"at Ulm, D-89069 Ulm, Germany
}
\date{\today}
\begin{abstract}
A procedure is discussed for creating coherent superpositions of motional states of ion strings. The motional states are across the structural transition linear-zigzag, and their coherent superposition is achieved by means of spin-dependent forces, such that a coherent superposition of the electronic states of one ion evolves into an entangled state between the chain's internal and external degrees of freedom. It is shown that the creation of such an entangled state can be revealed by performing Ramsey interferometry with one ion of the chain. 
\end{abstract}
\maketitle

\section{Introduction}

The superposition principle is one of the fundamental elements of quantum mechanics, most fascinating as it defies the everyday experience of physical reality. A paradigmatic example is Schr\"odinger's {\it Gedankenexperiment}, which pushes the quantum-mechanical predictions to their ultimate consequences, by envisioning an experiment that would lead to a quantum superposition involving a photon and a cat \cite{Schroedinger_1935_all,Zurek_1991, Bergou_Englert_1998}. The ``cat paradox'' has been extensively discussed in the literature. The realization of quantum superpositions of physical objects having size bigger than a single atom or photon has been experimentally pursued in several milestone experiments, some of which are reported in Refs. \cite{Haroche_2003,Leibfried-et-al-2003,Haeffner-Roos-Blatt-2008,Weinfurter_2009,Zeilinger_2009,Arndt_2011,Cleland_2010}. These efforts ultimately aim at controlling the quantum dynamics of increasingly large systems, applying mostly a bottom-up approach. The goal is to warrant scalability of quantum mechanical operations, a basic requisite for quantum technological applications.

Strings of trapped ions are one of the most promising platforms for quantum technologies. Among several remarkable achievements, we mention here that the collective vibrations of small arrays of ions have been cooled to the zero-point energy \cite{Jost_et_al_2009}, quantum teleportation of an ion's internal state has been realized \cite{Blatt_2004,Wineland_2004,Olmschenk_et_al_2009}, entangled states of the internal excitations have been demonstrated \cite{Leibfried-2006,Monz_et_al_2011}. First realizations of quantum simulators with these systems have been reported most recently \cite{Johanning_Varon_Wunderlich_2009, Islam_et_al_2011, Lanyon_et_al_2011, Schneider_Porras_Schaetz_2011,Blatt_2011}. This progress is mostly based on scaling up techniques, whose efficiency has been demonstrated for few particles, and points towards the need of identifying novel tools, which can allow one to control the quantum dynamics of mesoscopic systems.

\begin{figure}[bt]
\includegraphics[width=0.45\textwidth]{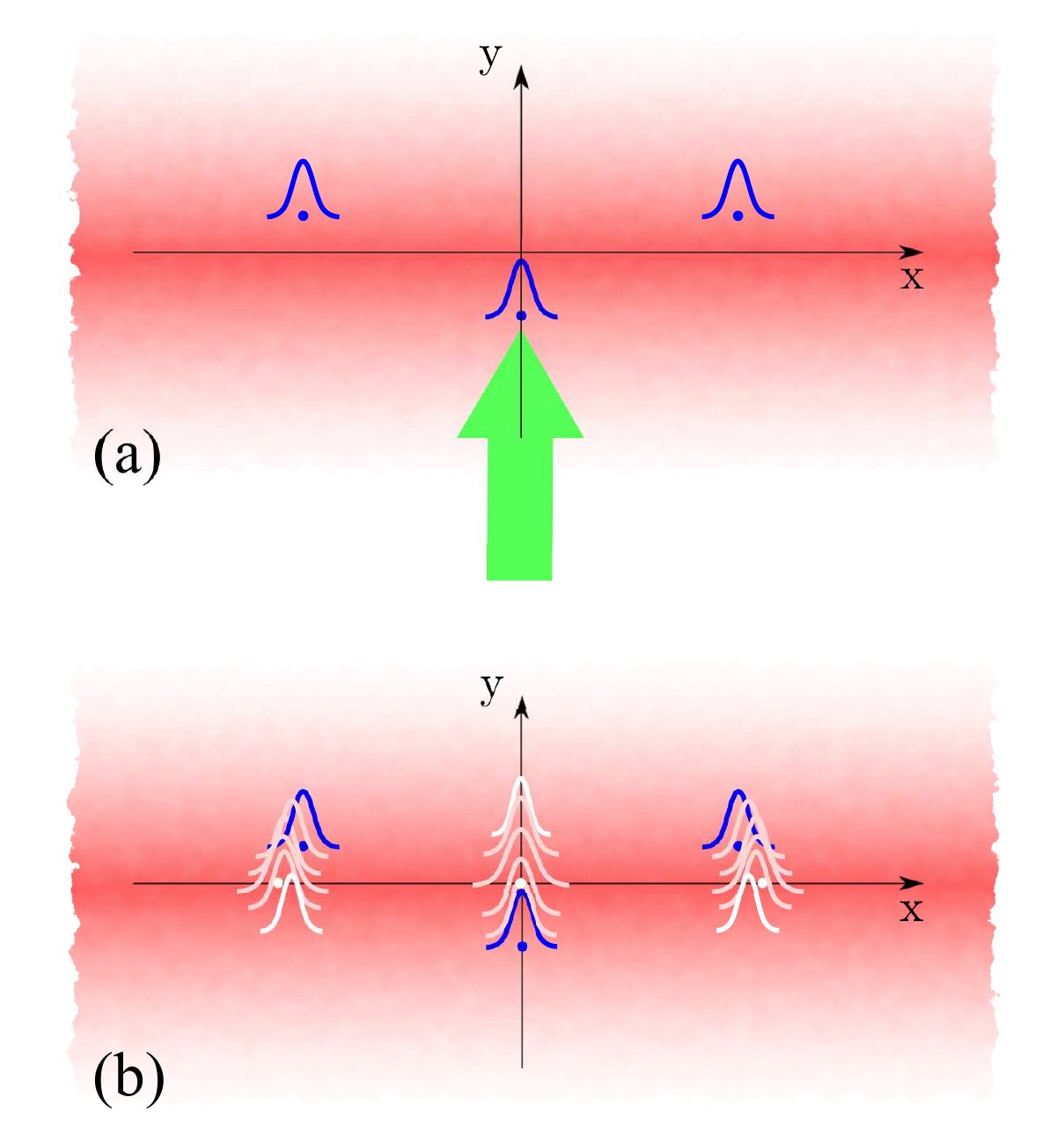}
\caption{\label{f:singleexcitation}(color online) \; Scheme for the creation of coherent superpositions of different motional states. (a)~The ions are initially in the internal state $ \ket g $ and the collective motion is in the ground state of a zigzag structure. A resonant laser pulse (green arrow) prepares the central ion in the superposition $\frac{1}{\sqrt{2}}\left( \ket g + \ket e \right)$. (b)~A state-dependent potential, acting only when the ion is in state $|e\rangle$, induces a conditional dynamics such that the ions internal and external degrees of freedom get entangled. When the state-dependent potential is sufficiently tight, the excited component will start oscillating around the equilibrium positions of the linear chain.
}
\end{figure}

In this article we propose a scheme for creating quantum superpositions of two different structures of a trapped ion crystal: the linear and the zigzag order \cite{Birkl-Kassner-Walther-1992}. The superposition is accessed by driving the electronic transition of one ion in the chain, in a setup where an external field makes the trap frequency spin-dependent. The setup is sketched in Fig.~\ref{f:singleexcitation}. The corresponding stability diagram is derived for three ions and a measurement scheme, to reveal the presence of entanglement, is discussed, which is based on Ramsey interferometry. 

Before we start, we note that the structural transition from a linear to a zigzag chain has been extensively studied in the literature \cite{MorigiFishmanPRL2004, Fishman-et-al-2008, Retzker_et_al_2008, Shimshoni_Morigi_Fishman_2011}. In Ref.~\cite{Fishman-et-al-2008} it was shown that in the thermodynamic limit this is a second order phase transition, which is classically described by the Landau model and whose control fields are the transverse trap frequency and the interparticle distance. If the system is quenched, the density of defects is well reproduced by the Kibble-Zurek model~\cite{Kibble_1976, Zurek_1985, delCampo_et_al_2010, De_Chiara_et_al_2010}. Our study is a first step towards creating a quantum superposition of two phases across the quantum phase transition, while the quenching here is realized by coupling to a quantum degree of freedom: a two-level transition of one ion in the chain. In this respect, our proposal extends to the ion-trap setup previous studies which considered this idea for spin chains coupled to a qubit~\cite{Quan-et-al-2006, Zhang_et_al_2008, Cormick_Paz_2008}. 

This article is organized as follows. In Sec.~\ref{sec:model} we introduce the setup and the corresponding model. In Section~\ref{sec:superpositions} schemes for preparing quantum superpositions of crystalline structures are proposed, while the details on how state-dependent crystalline structures can be created are reported in Sec.~\ref{sec:potentials}. Measurement protocols for detecting the quantum superposition are discussed in Sec.~\ref{sec:overlap}. The conclusions are drawn in Sec.~\ref{sec:conclusion}, while the Appendix provides further details on the stability diagrams. 

\section{Theoretical model}
\label{sec:model}

We consider $N$ atomic ions of mass $m$ and charge $q$, which are confined by an anisotropic harmonic potential (generated by a linear Paul trap \cite{Ghosh-1995} or a linearized Penning trap \cite{Powell-et-al-2002}). We denote by $\vek{r}_j$ and $\vek{p}_j$ the position and canonical conjugate momentum of the $j$-th ion, with $j=1,\ldots, N$. The internal degrees of freedom of the ions, which are relevant to the dynamics, are the electronic states $|g\rangle$ and $|e\rangle$, which we assume to be stable. They are driven by external laser pulses, which manipulate the quantum state of each ion's two-level transition in a controlled way. 

The total Hamiltonian $H$ describing the dynamics of the ions' internal and external degrees of freedom (excluding the laser pulses) can be decomposed into the sum
\beq \label{eq:Hamiltonian1}
H = H_{\rm at}+H_{\mathrm{kin}}  + V \,,
\eeq
where $H_{\rm at}=\sum_{j=1}^N\hbar\omega_0|e\rangle_j\langle e|$ and $H_{\mathrm{kin}} = \sum_{j=1}^N\vek{p}_j^2/(2m)$ are the internal and the kinetic energy of the ions, respectively, while $V$ is the mechanical potential. The latter takes the form 
\begin{equation}
V=V_{\mathrm{Coul}}+ V_{\mathrm{pot}}\,,
\end{equation}
with 
\begin{equation}
V_{\mathrm{Coul}} = \frac{1}{2}\sum_{j=1}^N \sum_{\substack{k=1 \\ k\neq j}}^N \frac{q^2}{4\pi \epsilon_0} \frac{1}{|\vek r_j -\vek r_k|} 
\end{equation}
the unscreened Coulomb repulsion, with $\epsilon_0$ the vacuum permittivity. The term $V_{\mathrm{pot}}$ denotes an external, harmonic potential, that is assumed to be anisotropic. In the rest of this work we take that the confinement along the $z$-direction is so tight, that the motion along this direction can be considered to be frozen out. In the $x$--$y$ plane, the potential depends on the internal state of the ions and takes the form 
\begin{equation} \label{eq:external potential}
V_{\mathrm{pot}} = \sum_{j=1}^N   \ketbraIX{g}{j}{g} \,V_g(\vek r_j) +  \ketbraIX{e}{j}{e} \, V_e(\vek r_j) \,,
\end{equation}
where 
\begin{align}
V_g(\vek r_j) &= \frac{1}{2} m \nu_x^2 \left( x_j^2 + \alpha^2 y_j^2 \,\right),  \\ 
V_e(\vek r_j) &= \frac{1}{2} m \nu_x^2 \left[ x_j^2 + (\alpha+\delta\alpha)^2 y_j^2 \,\right].
\end{align}
Here $\nu_x$ is the trap frequency along the $x$-axis,  $\alpha=\nu_y/\nu_x$ is the aspect ratio between transverse and axial trap frequency when the ion is in the internal state $\ket{g}$, and  
$ \alpha+\delta\alpha$ is the aspect ratio when the ion internal state is $\ket{e}$.

The potential (\ref{eq:external potential}) can be obtained by superimposing an optical potential to the ion trap. Such potential can be generated by a laser which is far detuned from a dipole transition between the state $\ket{e}$ and an additional auxiliary electronic level, while it does not affect the dynamics when the ion is in state $\ket{g}$ \cite{Grimm_W_O_1999,Schneider-et-al-2010}. Alternatively, the state-dependent potential can be realized by means of magnetic fields, by appropriately coupling a magnetic dipole transition \cite{Johanning_Varon_Wunderlich_2009}.

In this article we assume that the ions are sufficiently cold, such that they are localized around the equilibrium positions of the total potential, composed of $V_{\mathrm{Coul}}+V_{\mathrm{pot}}$.  The ordered structures they form depend on the the potential, and thus also on the ions' internal state. The potentials we will discuss support the linear and zigzag structures, which are illustrated on the top of Fig. \ref{f:eigenfrequencies}. 

For later convenience, we introduce dimensionless variables: the lengths are scaled by a characteristic scale $$l=q^{2/3} /( 4\pi\epsilon_0m\nu_x^2)^{1/3}\,,$$
which is the typical interparticle distance along the chain axis, while the energies by the scale $\mathcal E = m \nu_x^2 l^2$. The dimensionless potentials read
\begin{align}
& V_{\mathrm{Coul}} = \frac{1}{2}\sum_{j=1}^N \sum_{\substack{k=1 \\ k\neq j}}^N \frac{1}{|\vek r_j -\vek r_k|},\\
& V_g(\vek r_j) = \frac{1}{2} \left( x_j^2 + \alpha^2 y_j^2 \,\right), \\
& V_e(\vek r_j) = \frac{1}{2} \left[ x_j^2 + (\alpha+\delta\alpha)^2 y_j^2 \,\right]\,,
\end{align}
where for simplicity we keep the same notation for dimensionless variables as the one we used before the rescaling.

\section{Cat states of ion chains}
\label{sec:superpositions}

In this section we describe a possible scheme for preparing and measuring a coherent superposition of different motional states by manipulating the ions with lasers. We are interested in the case in which the chain is composed by few ions, for instance, three. We assume that the superimposed optical potential is sufficiently tight such that, when an ion in the zigzag chain is excited, the structure becomes  linear. In this regime the cat state is achieved by preparing a single ion in a superposition state.  The parameters required in order to access this regime will be analysed in the next section. 

\subsection{Creation of cat states}

The initial state of the crystal is assumed to be, unless otherwise stated, given by all ions in their electronic ground state  and the chain in the ground state of the zigzag structure, which we denote by $|\phi(0)\rangle=\ket{0}_{\mathrm{zz}}$. A laser pulse addresses one of the ions in the chain and drives the two-level transition $|g\rangle\to |e\rangle$,  performing the rotation $|g\rangle\to |g\rangle \cos\Theta/2+|e\rangle\sin\Theta/2$. Assuming that the duration of the pulse is much shorter than the typical scales of the vibrational motion, the state of crystal after the pulse is $|\Psi\rangle=|\psi_I\rangle\ket{0}_{\mathrm{zz}}$, with (for $\Theta=\pi/2$)
\beq	\label{eq:singelioninternalsuperposition}
|\psi_I\rangle= \frac{1}{\sqrt 2} \Big(\ket{ggg} + \ket{geg} \Big)  \,.
\eeq
After the pulse, the motional and internal degrees of freedom get entangled by the unitary evolution governed by Hamiltonian $H$, Eq. (\ref{eq:Hamiltonian1}), creating states of the form
\begin{equation} \label{eq:superposition-time evolved}
|\Psi(t)\rangle=\frac{ 1 }{\sqrt 2} \Big( \ket{ggg} \ket{0}_{\mathrm{zz}} + \ket{geg} |\phi(t)\rangle\Big) \,,
\end{equation}
with
\begin{equation}
|\phi(t)\rangle=e^{-iH_et/\hbar} \ket{0}_{\mathrm{zz}} 
\end{equation}
and $H_e=\langle e|H|e \rangle$ the Hamiltonian projected over the excited state. Here, the energy of state $\ket{g} \ket{0}_{\mathrm{zz}}$ is set equal to zero, and the state is in the reference frame of Hamiltonian $H_{\rm at}$.  Entanglement between internal and external degrees of freedom is achieved for times over which the overlap 
\begin{equation}
I(t)=\langle \phi(0)|\phi(t)\rangle
\end{equation}
has modulus smaller than unity. This entanglement can also be interpreted in terms of which-way information. Entanglement between internal and external degrees of freedom, in fact, diminuishes the visibility of the Ramsey signal introducing a ``distinguishability'' (which-way information) in the interferometric path. Maximal distinguishability, and correspondingly zero visibility, is found for $I=0$. We refer the reader to Refs. \cite{Englert_PRL,DeChiaraPRA2008} for a more detailed discussion.

\subsection{Ramsey interferometry}

The overlap $I(t)$ can be measured by means of Ramsey interferometry, according to the scheme proposed in Ref. \cite{DeChiaraPRA2008} and in~\cite{Pruttivarasin_et_al_2011} to study spin-dependent dynamics in ion chains. Let us assume that, after obtaining state (\ref{eq:superposition-time evolved}) for a given evolution time $t$, we apply a $-\pi/2$ pulse, i. e. the inverse operation of a $\pi/2$ pulse. The resulting state reads
\begin{equation} \label{eq:superposition-Ramsey}
|\psi_{II}(t)\rangle=\frac{1}{2} \Big[ \ket{ggg} (\ket{\phi(0)}+\ket{\phi(t)}) - \ket{geg} (\ket{\phi(0)}-\ket{\phi(t)}) \Big] \,,
\end{equation}
and the corresponding probability to find the addressed ion in state $\ket{g}$ is given by
\begin{equation}
{\rm P}_1(g) = \frac{1 + {\rm Re}\{I(t)\}}{2}.
\end{equation}
Alternatively, after obtaining the state (\ref{eq:superposition-time evolved}), we can first apply a phase gate on the addressed ion, namely an operation that maps the internal states according to the rule:  $\ket{g} \to \ket{g}$, $\ket{e} \to - i\ket{e}$. After an elapsed time $t$, the $-\pi/2$ pulse is applied and the probability to find the ion in $\ket{g}$ reads
\begin{equation}
{\rm P}_2(g) = \frac{1 + {\rm Im}\{I(t)\}}{2}.
\end{equation}
From these two measurements the overlap $I(t)$ is obtained.
We note that this quantity is sometimes called ``Loschmidt echo'', or also ``quantum fidelity'' and has been studied extensively in other systems, for instance in Refs. \cite{Quan-et-al-2006, Jacquod-Silvestrov-Beenakker-2001, Cucchietti-et-al-2003, GorinPhysRep2006}.

\subsection{Discussion}

Some remarks are now in order. First of all, in the example provided here it is the ion in the middle of the chain which is addressed by the laser pulses. The reason for this choice is rather natural, as the type of excitation does preserve the symmetry by reflection about the center of the trap, which characterises both linear and zigzag chain. Control of the internal excitation is a necessary condition for the implementation of the protocol. The selective driving of the central ion could be implemented by a focused laser beam, as realised for instance in Ref.~\cite{Blatt_2004}, or by means of a magnetic field gradient, which tunes only the central ion's transition in resonance with an external microwave field~\cite{Balzer_et_al_2006}. In this regime mechanical effects of the coupling to radiation can be taken to be very small and can be neglected\footnote{Neglecting the mechanical effects of the laser pulse simplifies the theoretical treatment but does not limit the feasibility of the protocol.}. The Hamiltonian for the pulse reads
\beq \label{eq:Hpulse}
H_L = \frac{i\hbar\Omega}{2} (\ketbra{g}{e} - \ketbra{e}{g}) \theta(t)\theta(t_{\rm pulse}-t)
\eeq
in the reference frame rotating at the laser frequency; $\theta(t)$ is the Heaviside function, describing the step-like form of the pulse with duration $t_{\rm pulse}$. Here, $\Omega$ is the Rabi frequency, and the rotation angle of the dipole is $\Theta/2=\Omega t_{\rm pulse}$.  

Neglecting the evolution of the vibrational motion during the pulse is justified provided that the largest vibrational frequency scale, $\nu_{\rm max}$, is such that $\nu_{\rm max}t_{\rm pulse}\ll 1$. Considering that $\Omega t_{\rm pulse}=\pi/4$ for the example here considered, the requirement can be rewritten in terms of the inequality $\nu_{\rm max}\ll \Omega$. 

We note that the scheme here proposed is based on creating a local defect when the central ion is in the excited state: this defect will induce another ordering (e.g., when starting from a zigzag array, the ion excitation induces a transition to a linear chain) provided that the length of the string is shorter than the correlation length~\cite{Fishman-et-al-2008}. However, as the number of ions is scaled up, a linear superposition like the state in Eq.~(\ref{eq:singelioninternalsuperposition}) will present a central deformation of finite size. This problem could be solved by creating a  Greenberger-Horne-Zeilinger (GHZ) state, namely,
\beq
 \frac{1}{\sqrt 2} \left( \ket{ggg} + \ket{eee}  \right) \,,
\eeq
which warrants that all ions feel the same potential. This kind of state has been realized in~\cite{RoosScience2004,LeibfriedNature2005} \footnote{It should be noticed that its preparation usually makes use of collective excitations of the ion chain, such that the free evolution of the vibrational motion during the preparation pulses cannot be neglected. Nevertheless, this just implies that the initial state of the chain is not the ground state of the motion but still a pure state}. The preparation of this class of states becomes increasingly difficult for larger ion chains~\cite{RoosScience2004,LeibfriedNature2005}, and the entanglement of GHZ-states is in general not robust \cite{Hein_2005}. 

To conclude this section, we mention that an alternative procedure for generating a cat state could be implemented by (i) first preparing the superposition state (\ref{eq:singelioninternalsuperposition}) for the internal degrees of freedom, and then (ii) slowly switching on the additional state-dependent potential, such that the vibrational state follows adiabatically the change in the  (state-dependent) vibrational Hamiltonian. In this form, it would be possible to prepare a superposition of two different structures in their respective ground state. This proposal poses several challenges and it will be elaborated in more detail in a subsequent publication.

\section{State-dependent crystalline structures}
\label{sec:potentials}

In this section we shall analyse the equilibrium configuration and normal modes of the ion chain in the harmonic trapping potential. The stationary equilibrium positions $\vek r_j^{(0)}$ satisfy the equations $\frac{\partial V}{\partial r_j} (\vek r_j^{(0)})= 0$, and they are stable when the eigenvalues of the Hessian matrix of $V$ evaluated at $\vek r_j^{(0)}$ are larger than zero. In this case the eigenvalues give the frequencies of the normal modes and the corresponding eigenvectors are the eigenmodes \cite{Goldstein}. Clearly, in our case there will be different sets of solutions depending on the ions' internal state.

In the following we will consider that the ion excitation couples the ground states of the {\it classical phase transition} \cite{Fishman-et-al-2008}. We will hence discard the regime, in which one would observe the disordered phase, predicted for the quantum phase transition, and which is observed when the trap frequency is close but below the critical value, yet quantum fluctuations destroy the zigzag order \cite{Shimshoni_Morigi_Fishman_2011,Shimshoni_Morigi_Fishman_PRA_2011}. This would correspond to a narrow interval of values, that has been characterised in detail in Ref. \cite{Shimshoni_Morigi_Fishman_PRA_2011}. In Ref. \cite{Retzker_et_al_2008} a characterisation of the parameters required for small chains of ions can be found.

\subsection{Spatial organization of the ions in harmonic potentials}
\label{sec:potentials:subsec:homogeneous}
We first examine the equilibrium structures for three trapped ions in the case in which the internal state of all particles is $\ket{g}$ and  the confinement in the $y$-direction is tighter than that along $x$, i.e., $\alpha>1$. The following discussion reviews results which have been reported, for instance, in \cite{James_1998,MorigiFishmanPRE2004,Fishman-et-al-2008}. For this case there are two different possible types of solutions for the equilibrium positions: the linear chain, with all ions aligned along the $x$-axis, and a planar structure, where the ions form the extremes of an isosceles triangle whose symmetry axis is aligned along the $y$-axis. We refer to this planar structure as zigzag configuration. More specifically, the equilibrium positions of the linear chain are 
\begin{align}
& x_1^{(0)} =-x_3^{(0)}=\sqrt[3]{5/4},\, x_2^{(0)} =0 ,  \label{eq:linear chain, x}\\
& y_1^{(0)}=y_2^{(0)}=y_3^{(0)}=0 .  \label{eq:linear chain, y}
\end{align}
This configuration is stable when $\alpha>\alpha_c$, with $\alpha_c=\sqrt{12/5}\approx 1.5492$. In this case, the axial normal mode frequencies are  $\{1, \sqrt 3, \sqrt{29/5}\}$ and the transverse ones are  $\{\sqrt{\alpha^2-\alpha_c^2},\sqrt{\alpha^2-1},\alpha \}$. 

\begin{figure}[hbt]
\begin{center}
\includegraphics[width=0.43\textwidth]{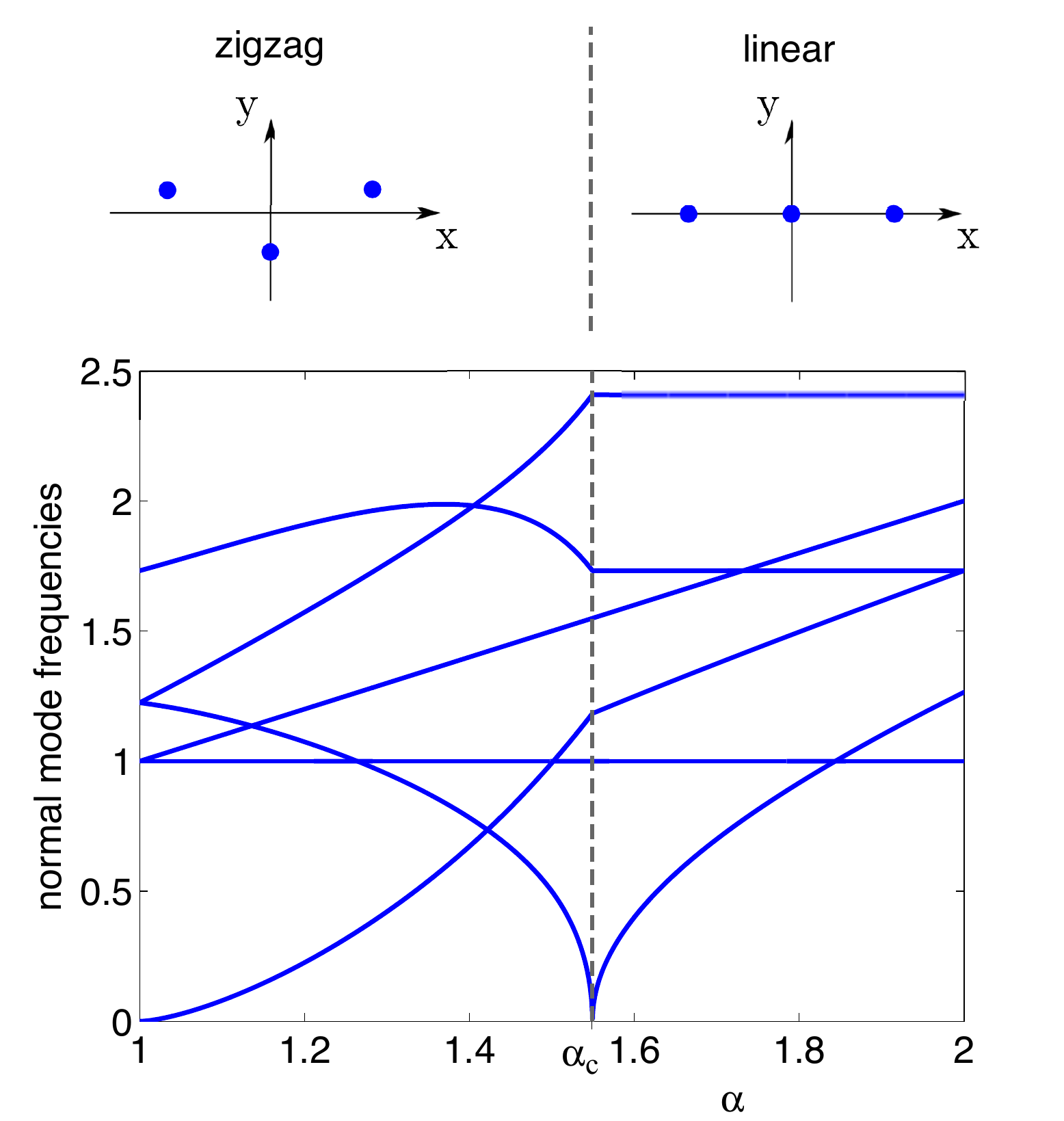}
\end{center}
\caption{\label{f:eigenfrequencies}
Normal mode frequencies for three ions confined in a harmonic potential, in units of the trap frequency $\nu_x$, as a function of the aspect ratio $\alpha = \nu_y/\nu_x$. The dashed vertical line indicates the critical value $\alpha_{c}\approx 1.5492$, separating the zigzag ($\alpha<\alpha_{c}$) from the linear array ($\alpha>\alpha_{c}$).}
\end{figure}

For $\alpha<\alpha_c$ (and $\alpha>1$) the chain is in a zigzag configuration, with equilibrium positions \begin{align}
x_1^{(0)} &= -x_3^{(0)}=\bar x, & x_2^{(0)} &=          0,   \label{eq:eqposZZX}\\
y_1^{(0)} &= y_3^{(0)}= \bar y, & y_2^{(0)} &= -2 \bar y, 
\end{align}
where
\begin{align}
&  \bar x = \Big[4\Big(1-\frac{\alpha^2}{3}\Big)\Big]^{-1/3} \label{eq:eqposZZXxbar}\\
& \bar y = \pm \frac{1}{3}\Big[\Big(\frac{3}{\alpha^2}\Big)^{2/3}-\bar x^2\Big]^{1/2} \label{eq:eqposZZXybar} \,.
\end{align}
The analytic expressions for the normal mode frequencies are rather cumbersome and are reported, for instance, in Ref. \cite{Fishman-et-al-2008} for the case of $N$ ions in a ring. Figure \ref{f:eigenfrequencies} displays the frequencies of the normal modes as a function of the aspect ratio $\alpha$.

\subsection{Spatial organization in state-dependent harmonic potentials}
\label{sec:potentials:subsec:inhomogeneous}
We now extend the previous considerations to the case in which the internal state of the three ions is not the same, so that the ions experience different trapping potentials. We shall consider only the case in which one of the three ions is excited. In particular, we will discuss the case in which the excited ion is in the middle of a linear chain. The configurations found for the case, in which the excited ion is at one edge of the chain, are reported for completeness in the Appendix.

\begin{figure}[hbt]
\includegraphics[width=0.35\textwidth]{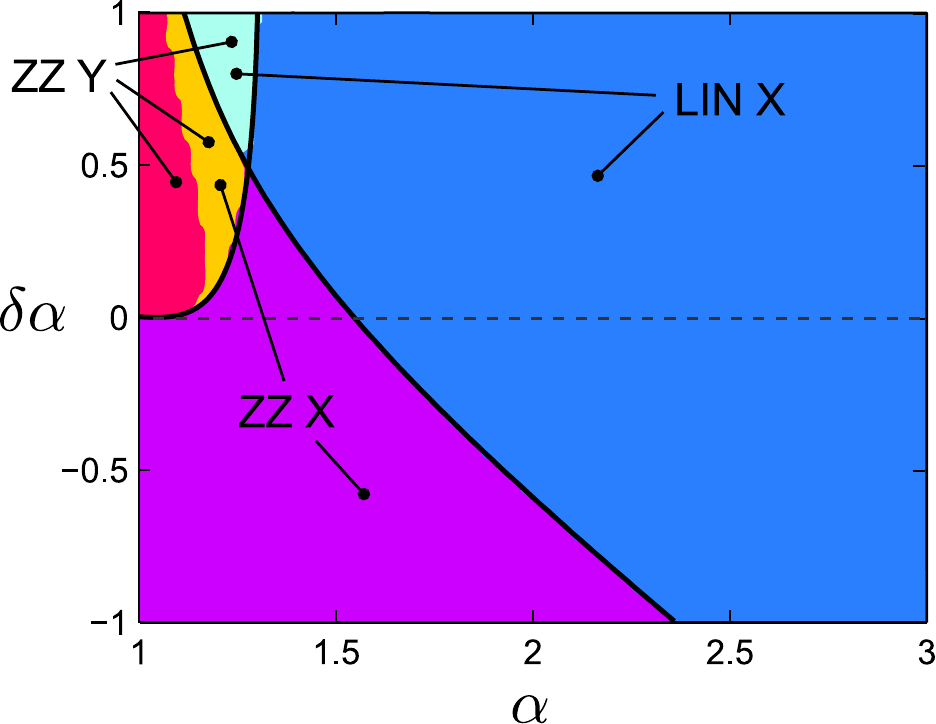}
\caption{
\label{f:central_ion_excited}
(color online) Stable configurations for a three-ion chain as a function of the aspect ratio $\alpha$ and variation~$\delta \alpha$ (here due to a state-dependent potential acting on the central ion). The labels ``LIN X'', ``ZZ X'' and ``ZZ Y'' correspond respectively to a linear chain along the $x$-axis,  a zigzag chain along the $x$-axis and a zigzag along the $y$-axis. The solid lines are the boundaries given in Eqs.~(\ref{eq:stability boundary 1}) and (\ref{eq:stability boundary 2}). The other boundary was determined numerically (the wiggled boundary is here due to the chosen numerical resolution). The variation of the state-dependent potential corresponds to a vertical displacement in the plot.
}
\end{figure}

When the ions form a linear array, the equilibrium configurations corresponding to state $|geg\rangle$ are the same as in the homogeneous case, and are thus given by Eqs.~(\ref{eq:linear chain, x}-\ref{eq:linear chain, y}). In this case, in fact, the central ion is located in the center of the trap where the external potential is zero. For this case the eigenfrequencies read
\begin{multline}
\left\{1,\sqrt{3},\sqrt{29/5},\sqrt{\alpha^2-1},\right. \\ 
\sqrt{\alpha ^2+\alpha  \delta \alpha -(12 - 5 \delta \alpha ^2 - \rho)/10 },\\
\left. \sqrt{\alpha ^2 + \alpha \delta \alpha -(12 - 5 \delta \alpha ^2 + \rho)/10 } ~\right\} \,, 
\end{multline}
with $\rho = \sqrt{128+[5\delta\alpha(2\alpha+\delta\alpha)-4]^2}$.  By imposing that the eigenfrequencies are real and positive, one finds that the linear chain is stable for $\delta\alpha>\delta\alpha_{\rm c}$, with
\begin{equation} \label{eq:stability boundary 1}
\delta\alpha_{\rm c} =  \left( 2 \sqrt{\frac{2}{5\alpha^2-4}} -1 \right) \alpha\,,
\end{equation}
where $\alpha \geq 1$. 
The corresponding curve is reported in the stability diagram of Fig.~\ref{f:central_ion_excited}. Here, one observes that for most values of $\delta\alpha<\delta\alpha_{c}$ the chain is in a zigzag structure along the $x$-axis (i.e. symmetric about the $y$-axis), whose equilibrium positions read
\begin{align}
x_1^{(0)} &= \bar x, & x_2^{(0)} &=          0, & x_3^{(0)}&= -\bar x, \\
y_1^{(0)} &=  \bar y, & y_2^{(0)} &= -2 R \bar y, & y_3^{(0)}&= \bar y \,,
\end{align}
with
\begin{align}
\bar x &= \Big[4\Big(1-\frac{\alpha^2}{1+2R}\Big)\Big]^{-1/3} \,,\label{eq:eqposZZXxbarInhom}\\
\bar y &= \pm \frac{1}{1+2R}\Big[\Big(\frac{1+2R}{\alpha^2}\Big)^{2/3}-\bar x^2\Big]^{1/2} \label{eq:eqposZZXybarInhom} \,,
\end{align}
and
\beq \label{eq:Rdef}
R=\frac{\alpha^2}{(\alpha+\delta\alpha)^2} \,.
\eeq
The homogeneous case is recovered in the formula by setting $\delta\alpha=0$: in this limit, $R=1$ and Eqs. \eqref{eq:eqposZZXxbarInhom}, \eqref{eq:eqposZZXybarInhom} coincide with Eqs. \eqref{eq:eqposZZXxbar}, \eqref{eq:eqposZZXybar}, respectively. 

The region  in the upper left corner of Fig.~\ref{f:central_ion_excited} corresponds to the appearance of zigzag configurations along the $y$-direction. The corresponding stability boundary is evaluated by imposing that the Hessian matrix, giving the normal modes, has vanishing determinant, which leads to the equation:
\begin{equation} \label{eq:stability boundary 2}
\frac{\alpha^2}{2R+1} = - \left(\frac{\bar y}{D}\right)^2 + \frac{1}{3} + \frac{\left(3\bar x \bar y\right)^2} {D^4 \left[2-\alpha^2-\left(\frac{\bar y}{D}\right)^2\right] } 
\end{equation}
where $\bar x$, $\bar y$ and $D$ correspond to the equilibrium positions in this configuration and are given by:
\begin{align}
& \bar x = \frac{1}{3}\Big(3^{2/3}-\bar y^2\Big)^{1/2} \label{eq:xeq-ZZY} \,, \\
& \bar y = \Big[4\Big(\alpha^2-\frac{1}{3}\Big)\Big]^{-1/3} \label{eq:yeq-ZZY} \,,\\
& D = \left(9 {\bar x}^2 + {\bar y}^2\right)^{1/2} = 3^{1/3}\,. 
\end{align}
We note that, as opposed to the case when all ions experience the same potential, more than one configuration can be stable in the same region.

\section{Loschmidt echo}
\label{sec:overlap}

A clear demonstration of the creation of cat states would require quantum tomography of the crystal state \cite{Deleglise_et_al_2008, Lutterbach_Davidovich_1997, Poyatos_et_al_1996, Leibfried-et-al-2003}. Nevertheless, signatures of the quantum superposition can be obtained  by means of Ramsey interferometry. This scheme allows one, as we have shown, to determine the overlap $I(t)$ as a function of the elapsed time $t$ between the two laser pulses. In this section we analyse the behaviour of the absolute value of the overlap $|I(t)|$ for different values of the axial and spin-dependent potential. This is performed by scanning through various regions of the phase diagram of the linear-zigzag structural instability for the case of three ions. For this calculation, the potentials were approximated by their quadratic expansions about the corresponding equilibrium positions, and the vibrations have been quantized according to the procedure reported for instance in Ref.~\cite{MorigiFishmanPRE2004}.

In the following discussion we will assume that the chain is initially either in the zigzag or in the linear configuration, depending on the value of the aspect ratio $\alpha$, and that the ions are prepared in state~\eqref{eq:superposition-time evolved} at $t=0$, as described in Sec.~\ref{sec:superpositions}. 
The strength of the harmonic potential trapping the central ion is given by the frequency $\nu_y'=\nu_y+\delta\nu_y$, with $\delta\nu_y=\delta\alpha\,\nu_x$.
\begin{figure}[p!hbt]
\includegraphics[width=0.4\textwidth]{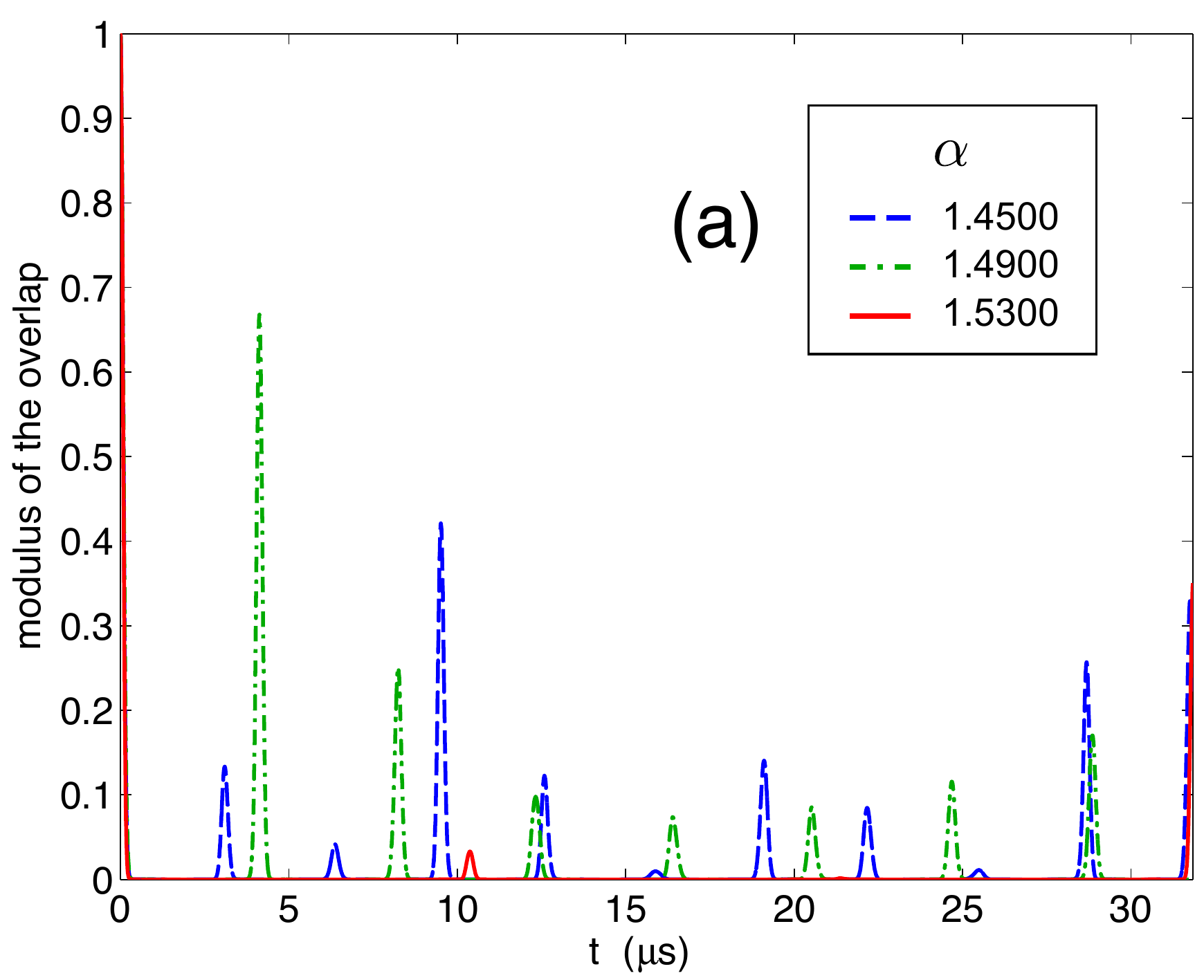} \\
\includegraphics[width=0.4\textwidth]{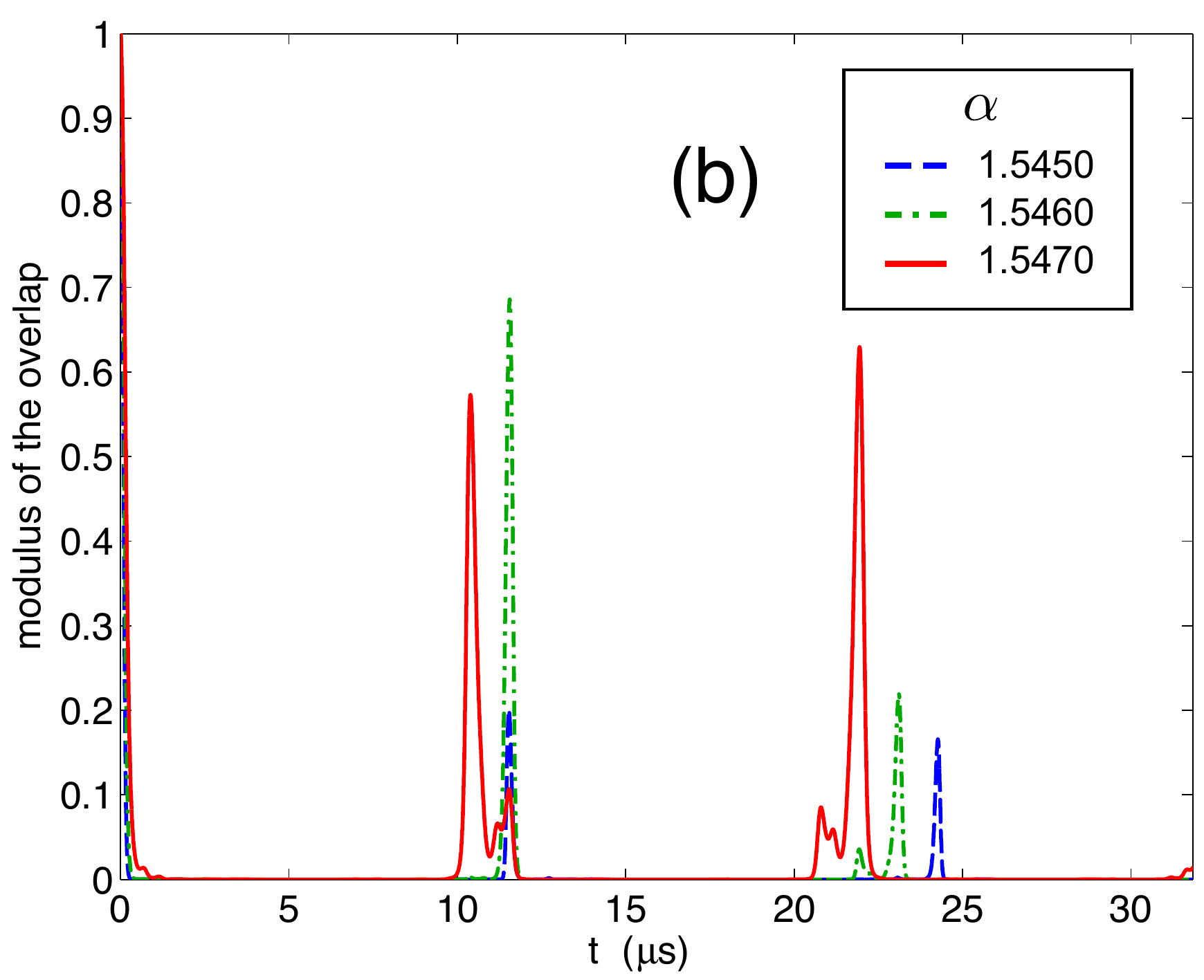} \\
\includegraphics[width=0.4\textwidth]{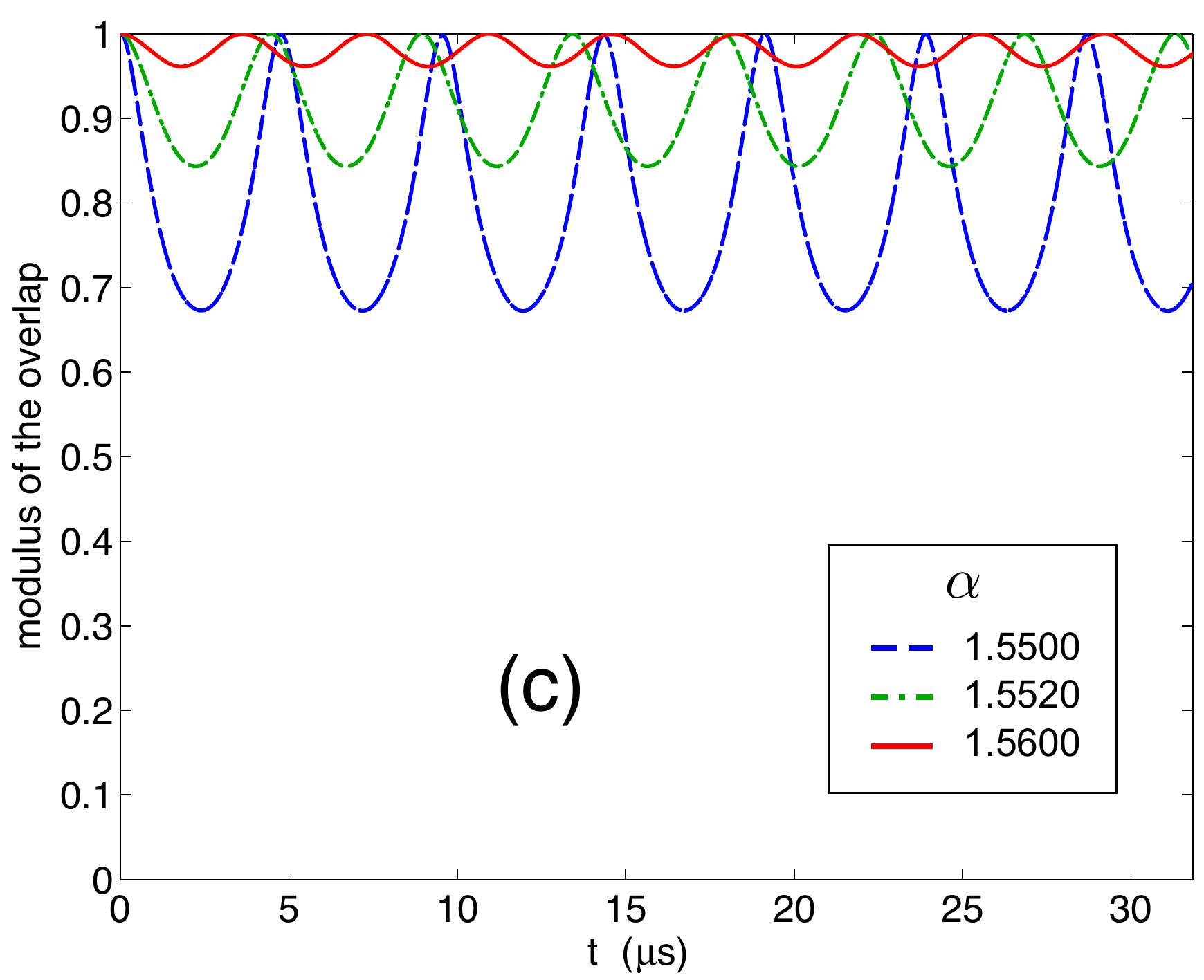}\\
\caption{\label{f:signals} Overlap signal $|I(t)|$ as a function of the elapsed time $t$ (in $\mu$s) for a Ramsey setup in which the central ion of a three-ion chain is excited. The curves are evaluated for $ {}^9\mathrm{Be}^+ $ ions for the parameters $\nu_x= 2\pi \times 500$~kHz, $\nu_y=\alpha\nu_x$, and $\nu_y'=\nu_y+ 2\pi\times 10 $~kHz (corresponding to $\delta\alpha =0.02 $). The values of $\alpha$ (and thus $\nu_y$) are given in the boxes. The equilibrium structures for ground and excited states are (a) zigzag-arrays with different transverse displacements (here $ \nu_y = 2\pi \times \{725, 745, 765\}$~kHz); (b) a zigzag- and a linear array, respectively ($ \nu_y = 2\pi \times \{772.5, 773.0, 773.5\}$~kHz); (c) linear arrays with different transverse confinements ($ \nu_y = 2\pi \times \{775.0, 776.0, 780.0\}$~kHz). 
} 
\end{figure}
We first focus on the signal for different values of $\alpha$, fixing the quench $\delta\alpha$. This corresponds to analysing the Ramsey signal by keeping constant the frequency shift $\delta\nu_y$ (due to the spin-dependent potential) but varying the potential at the electrodes. 

Figure~\ref{f:signals} displays $|I(t)|$ for different values of $\alpha$ when the excitation of the central ion gives rise to a transition (a)~between two zigzag structures (where the second one is more tightly confined), (b)~from a zigzag to a linear string, and (c)~between two linear strings (where the second one is more tightly confined). We note that the observed time-dependent behaviour is determined by the modes of the crystalline structure that is generated by exciting the central ion. In fact, the initial structure, which corresponds to the structure when all ions are in state $\ket{g}$, is the ground state of the corresponding array. In~(a) revivals of the overlap signal are observed. These revivals become more distant in time the closer the zigzag array is to the mechanical instability, where the time intervals scale with the period of the lowest eigenfrequency of the tighter zigzag chain. In~(b), where the excitation corresponds to a quench across the transition connecting zigzag and linear chain, one still observes revivals of the overlap, whereby the signal exhibits a more complex, quasi-periodic structure. When the transition is instead between two linear arrays with different transverse confinement~(c), the overlap exhibits a periodic modulation whose maximal amplitude is unity and at twice the frequency of the lowest frequency eigenmode: the zigzag mode of the linear chain. This modulation is a manifestation of squeezing induced by the sudden change of the trap potential.

\begin{figure}[p!hbt]
\includegraphics[width=0.4\textwidth]{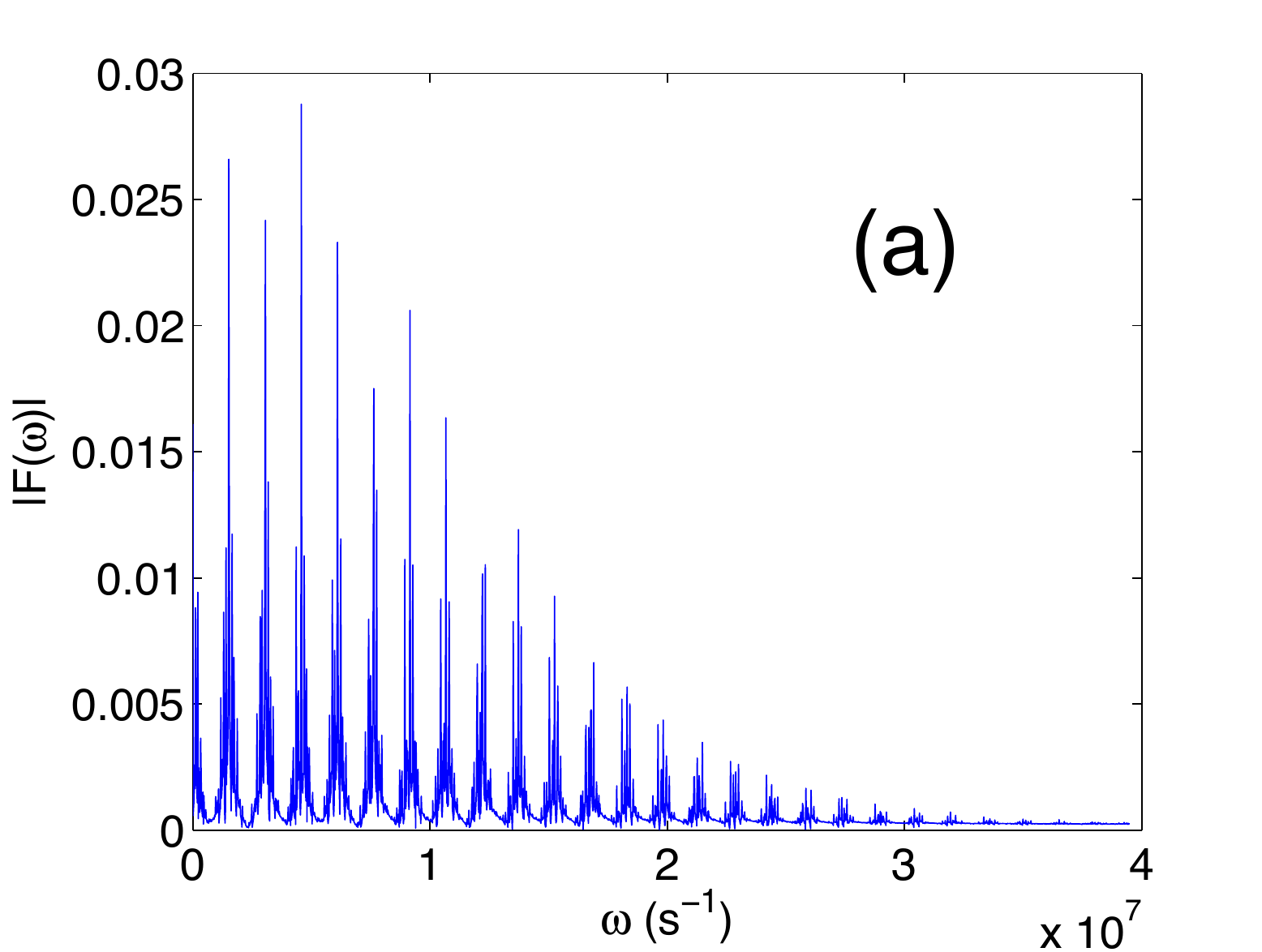} \\
\includegraphics[width=0.4\textwidth]{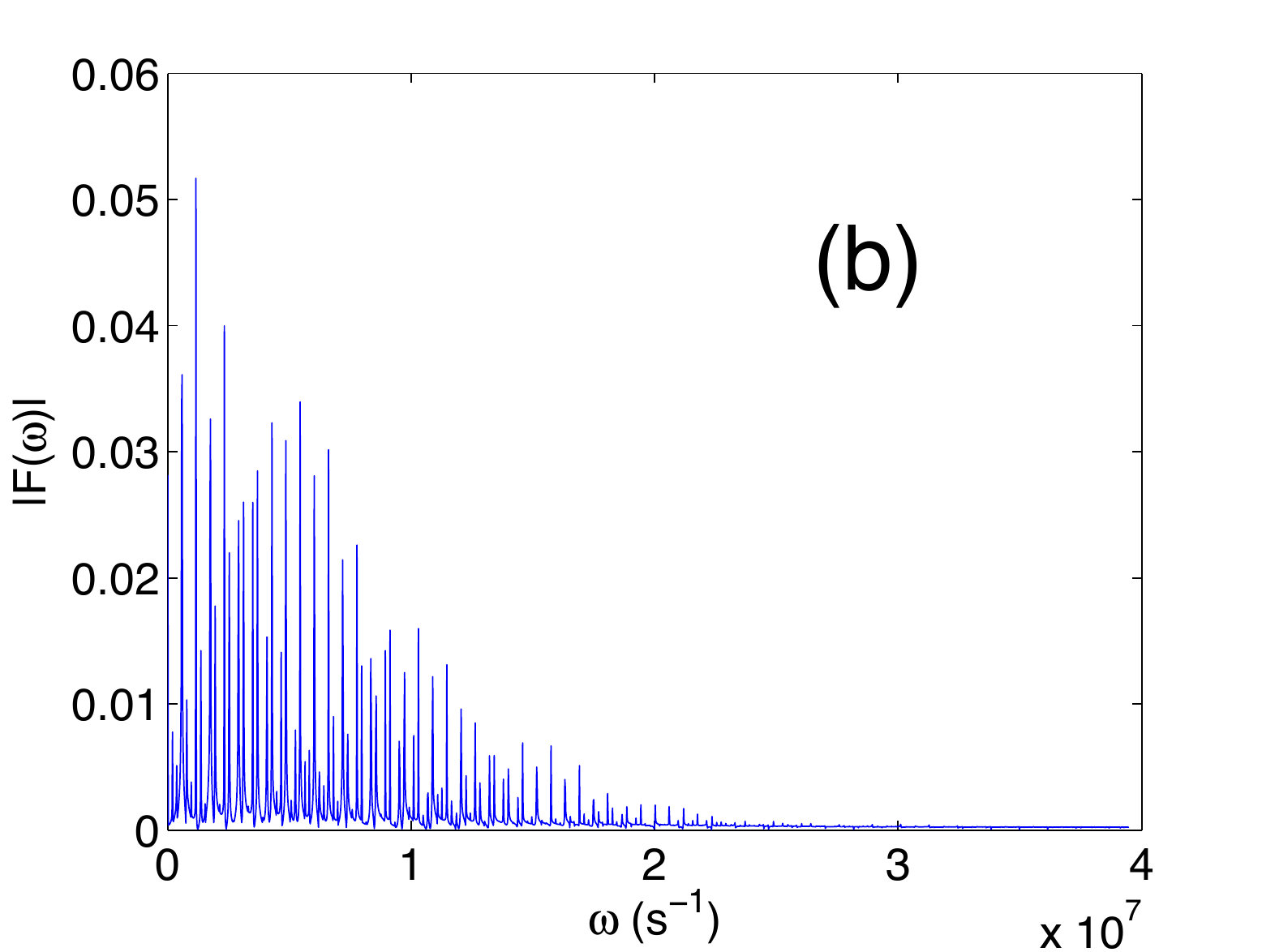} \\
\includegraphics[width=0.4\textwidth]{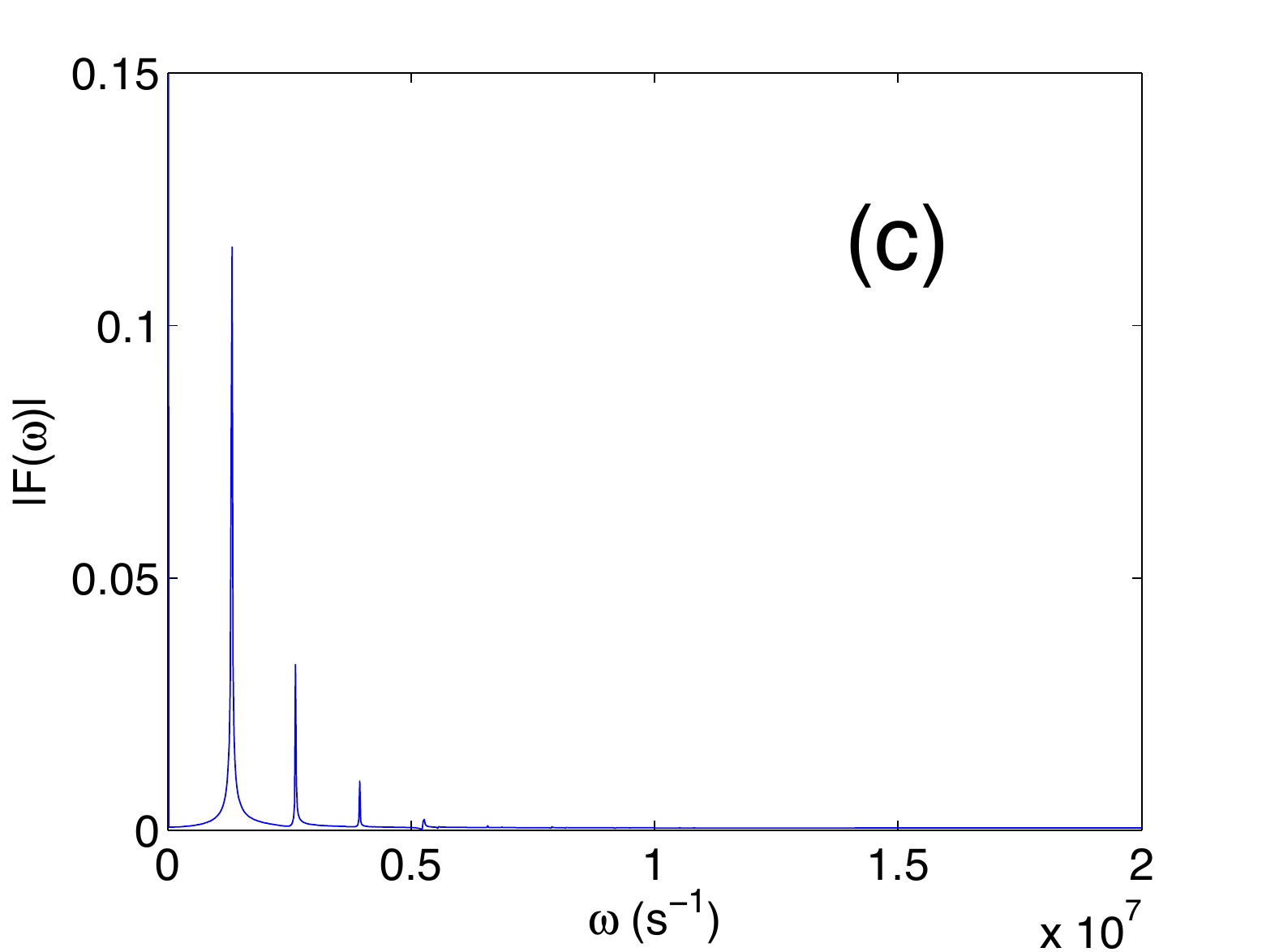}\\
\caption{\label{f:spectra} Fourier transform of the overlap signal $|I(t)|$ in Fig.~\ref{f:signals}. Here, (a) corresponds to the curve at $\alpha=1.49$, (b) to $\alpha=1.547$, and (c) to $\alpha=1.55$.
} 
\end{figure}

This behaviour can be further characterised using the Fourier spectra of the overlap signal. These are shown in Figs.~\ref{f:spectra} for each of the three cases displayed in Figs.~\ref{f:signals}. A first look shows that the spectral components of the transition between the zigzag structures are dominated by a comb of frequencies, Fig.~\ref{f:signals}(a), with a background signal. The comb is at frequencies that are multiple of the lowest frequency of the tight zigzag array. This structure splits up into more components when the transition couples two different structures, Fig.~\ref{f:signals}(b). Finally, when the transition couples two linear structures, Fig.~\ref{f:signals}(c), the spectrum is much more regular and composed by only a few frequency components, dominated by the lowest one. Further analysis shows that the dominant contribution in Fig.~\ref{f:signals}(a) and (c) is from the zigzag mode, which is also the soft mode of the structural phase transition. This mode seems to play a relevant role also when the transition couples two structures across the structural instability, Fig.~\ref{f:signals}(b). The revivals are associated with the oscillations of this mode, which is mostly excited by the quench. 

We note that the appearance of the revival signal is a signature of the quantum superposition: this signal would be absent if there is no quantum coherence between the two crystalline states. We then focus on its features, and characterise them in detail considering electronic excitations that couple a crystal with an equilibrium zigzag structure to a crystal with a linear array.

\begin{figure}[p!hbt]
\includegraphics[width=0.4\textwidth]{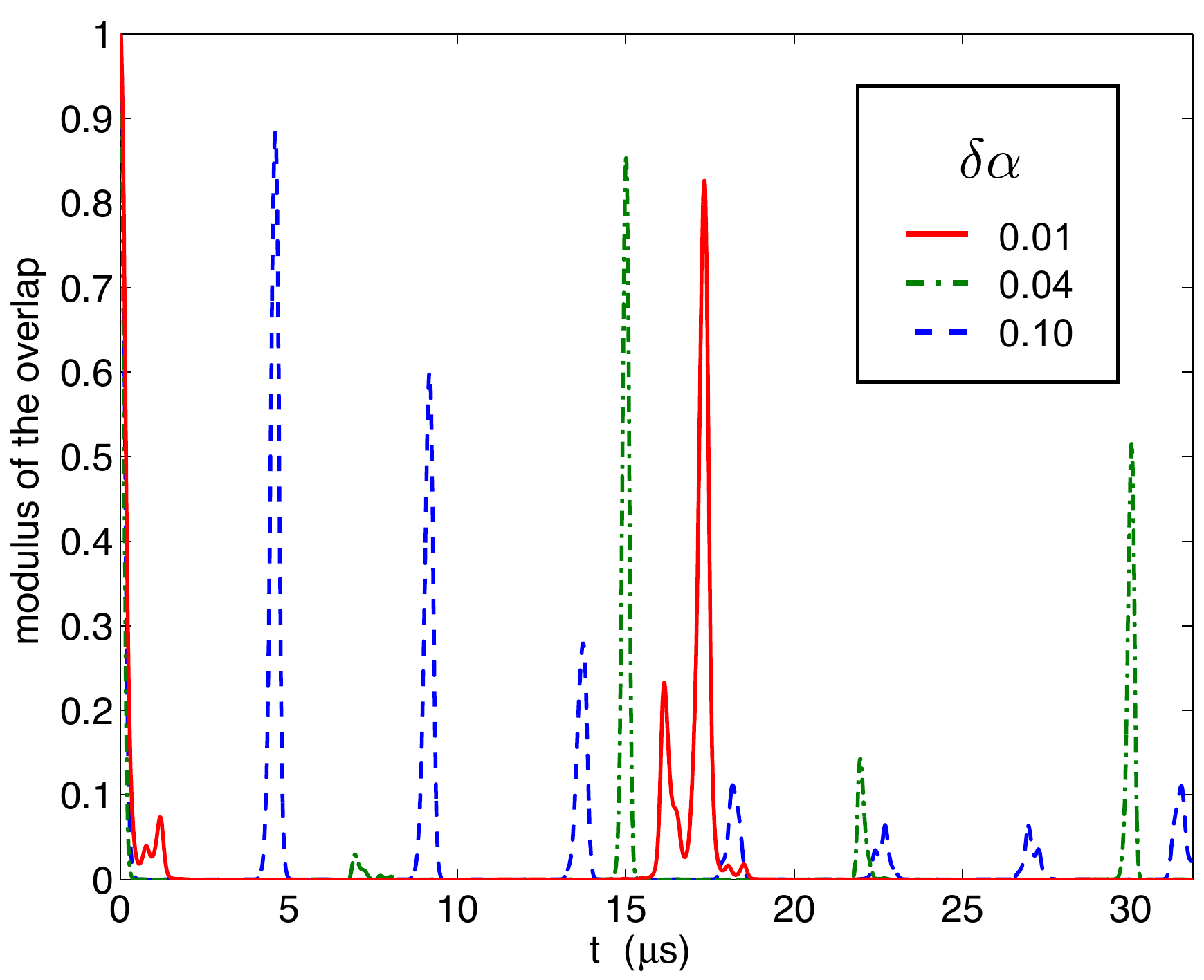}
\caption{\label{f:signals-2} Same as Fig.~\ref{f:signals} but for $\alpha=1.5470$ ($\nu_y = 2\pi \times 773.5 $~kHz) and  $\delta\alpha = \{0.01,0.04,0.10\}$ (corresponding to $\delta\nu_y = 2\pi\times\{5, 20, 50\} $~kHz). In this case the equilibrium structures for ground and excited states are a zigzag and a linear array, respectively.}
\end{figure}

\begin{figure}[p!hbt]
\includegraphics[width=.4\textwidth]{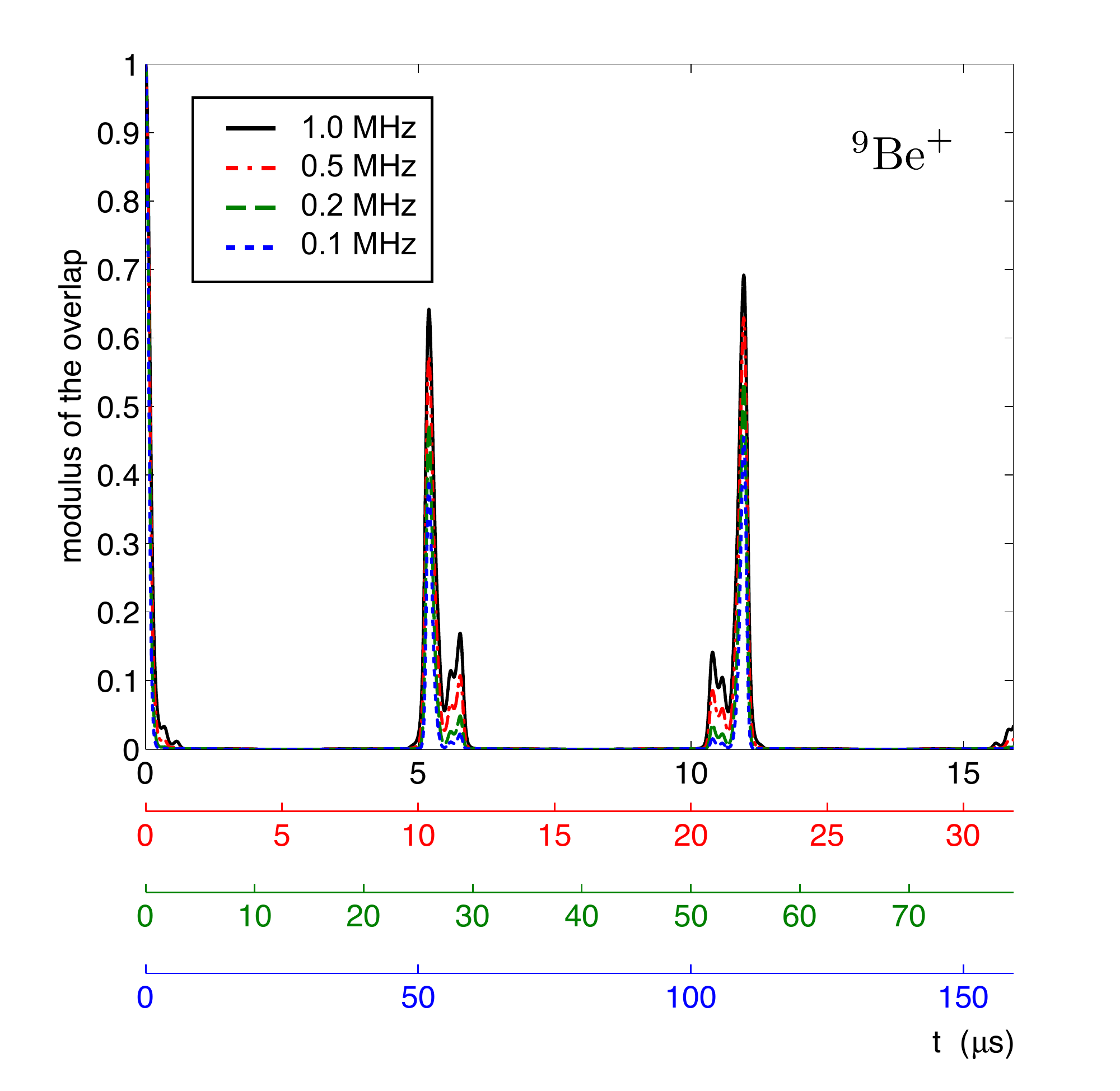} 
\caption{\label{f:signals-3} Same as Fig.~\ref{f:signals-2} but for $\delta\alpha=0.02$ and different axial trap frequencies $ \nu_x $ (the values are reported in the box). The value for $ \alpha $ corresponds to  $ \nu_y = 2\pi \times \{154.7,   309.4,    773.5,    1547\}$~kHz, the frequency shift due to the optical potential is $ \delta\nu_y = 2\pi\times \{2,4,10,20\} $~kHz (ordering corresponds to the values of $\nu_x$ reported in the legenda in the box, from bottom to top). The color-coded abscissa indicate the time (in $\mu$s) for the curves of the corresponding color.}
\end{figure}

The figures discussed in the following refer to the situations in which ground and excited states correspond to equilibrium structures across the mechanical instability, namely, when the initial state is a zigzag array and the excited state is a linear chain. Figure~\ref{f:signals-2} shows the overlap signal for different frequencies of the spin-dependent potential, corresponding to different values of the parameter $\delta\alpha$ (larger absolute values correspond to larger variations). The distance between the revival signals decreases as $\delta\alpha$ increases, namely, the excited state is deeper in the linear phase. Moreover, for $\delta\alpha=0.1$ one observes the onset of a periodic signal. The form of the signal at $\delta\alpha=0.02 $ is further characterised in Fig.~\ref{f:signals-3} as a function of the axial trap frequency, showing a slight increase of the revival signals as $\nu_x$ is increased, and correspondingly as the ions are more tightly confined in the $x$-direction. 

\begin{figure}[p!hbt]
\includegraphics[width=0.4\textwidth]{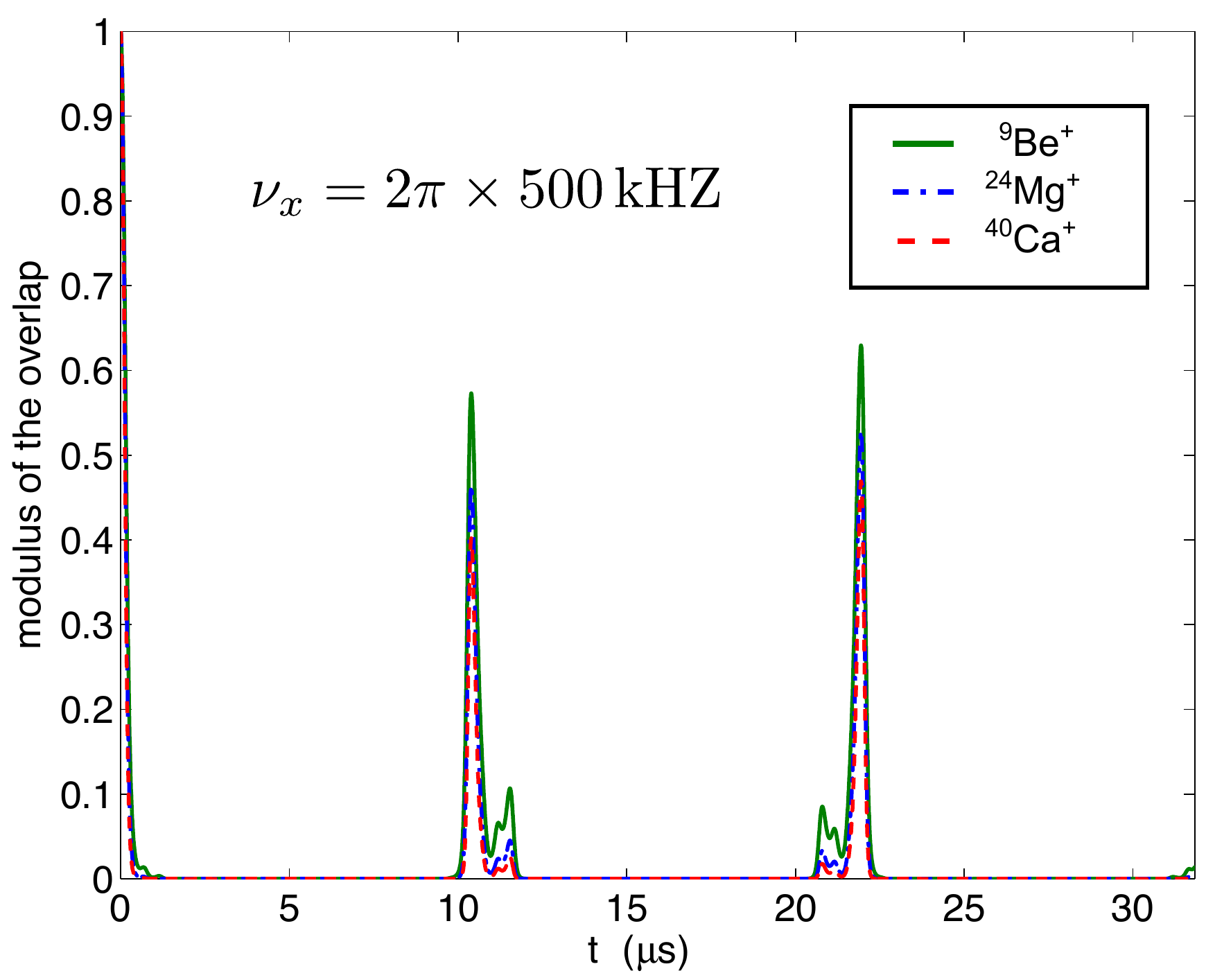}
\caption{\label{f:signals-comparison} Same as Fig.~\ref{f:signals-2} but for $\delta\alpha=0.02$ but for different ion species.} 
\end{figure}

Finally, the form of the revivals is reported in Fig.~\ref{f:signals-comparison} as a function of the ion species, that is, of the mass determining the spread of the motional wave packet. In this case, for lighter ions, i.e., for larger quantum fluctuations, one observes a slight increase in the revival signal.

The dynamics we considered so far is unitary: decay of the overlap signal is solely due to dephasing in the dynamics of internal and external degrees of freedom. Decoherence and noise sources have been neglected. These would affect both internal and external degrees of freedom. They are expected to introduce a damping factor in the overlap signal, setting an upper bound for partial revivals decreasing with time, such that a measured revival in an experiment in turn would give an estimated bound on decoherence effects. In order to observe the curves reported in Fig.~\ref{f:signals}, one needs coherence times longer than $100\nu_x^{-1}$ for both internal and external degrees of freedom. Taking $\nu_x\simeq 2\pi \times 500$~kHz, the state would need to remain coherent for times longer than 30~$\mu$s. This assumption seems reasonable given the coherence times reported for instance in~\cite{Deslauriers-et-al-2006, Haeffner-Roos-Blatt-2008, Timoney-et-al-2011}. The requirements of our proposal in terms of optical power are very moderate: for the values corresponding to Fig.~\ref{f:signals}, the optical potential should have an associated motional frequency of the order of $2\pi~\times$~100~kHz, well within present reach. Finally, we note that the ability to combine optical potentials and ion traps has already been demonstrated~\cite{Katori-Schlipf-Walther-1997, Schneider-et-al-2010}.

\section{Conclusions} \label{sec:conclusion}

In this paper we have studied the dynamics of the structural phase transition from the linear to the zigzag configuration in a chain of trapped ions, with the aim of creating a coherent superposition of different mesoscopic crystalline structures. Among several possibilities to achieve that goal, we have chosen to exploit spin-dependent trapping potentials, which influence each ion's motional state as a function of its internal electronic state. This allows in particular to create an entangled state from coherent superpositions of one degree of freedom.

Specifically, our protocol starts with an ion chain at rest under external confinement in a regime near the chain's structural phase transition. Excitation of one particular ion from the chain into a superposition of metastable electronic states affects, via the state-dependence of the external potential, this ion's motion, entangling its vibrational and internal state. As a result of Coulomb repulsion, the other ions in turn rearrange their motion according to this ion's new state, thereby becoming entangled with it as well. 

This realizes, under appropriate conditions described in the text,  a mesoscopic superposition (a cat state) of different crystalline structures, something that to our knowledge has not been observed or proposed before \footnote{After completion of this work we became aware of work along related lines carried out in:  W.~Li and I.~Lesanovsky, ArXiv e-print (2011), 1108.3591, where the creation of a three-ion state-dependent structure was proposed by means of Rydberg excitation of the central ion.}. Signatures of the generated cat state can be obtained by means of Ramsey interferometry, according to the scheme discussed in \cite{DeChiaraPRA2008}. Our analysis shows that these dynamics could in principle be observed in current experiments, using crystals of three ions.

Future work will focus on the properties of the Ramsey signal measured for longer chains when the ions are in the disordered phase of the quantum phase transition \cite{Shimshoni_Morigi_Fishman_2011,Shimshoni_Morigi_Fishman_PRA_2011}. Moreover, the efficiency of the protocol will be characterised as a function of the temperature of the chain, including also heating and decoherence effects systematically in the theoretical model. The control tools proposed here open the way to the application of optimal control techniques in the spirit of the proposals in Refs. \cite{Calarco1,Calarco2}, for the purpose of creating robust mesoscopic quantum states on demand.

\medskip

\section*{Acknowledgments}
The authors acknowledge fruitful discussions with Michael Drewsen, J\"urgen Eschner,  Endre Kajari, and Christof Wunderlich. This work was partially supported by the Spanish Ministry MICINN (QNLP FIS2007-66944 and ESF-EUROQUAM CMMC ``Cavity-Mediated Molecular Cooling''), the Alexander von Humboldt and the German Research Foundation, and the European Commission (STREP PICC, COST action IOTA, integrating project AQUTE).

\appendix

\section{Equilibrium configurations when one outer ion is excited} \label{subsubsec:outer ion}
We now determine the stability diagram when one of the outer ions is excited and experiences a state-dependent potential. We distinguish between three different cases: 1)~a linear chain configuration along the $x$-direction, which we denote by LIN~X*; 2)~a linear chain configuration along the $y$-axis, which we denote by LIN~Y*; and 3)~a triangular configuration which we label by TRIA* (here we choose not to distinguish between orientation along $x$ or $y$). The asterisk always denotes the structures where one of the outer ions is excited. As in the previous cases, we shall focus on the parameter region $\alpha\geq1$, for which a linear chain along the $y$-axis is never stable.

For the LIN~X* we obtain again the same equilibrium positions as in Eq.~(\ref{eq:linear chain, x}-\ref{eq:linear chain, y}). It is possible to determine the stability boundary analytically by solving equation
\beq \label{eq:stability boundary 3}
\frac{\alpha^2}{2R} = \frac{9}{20} + \frac{(65/8)\alpha^2-9}{5(5\alpha^4-25\alpha^2/2+4)} \,,
\eeq
where $R$ is defined as in Eq.~(\ref{eq:Rdef}).

\begin{figure}[htb]
\includegraphics[width=0.35\textwidth]{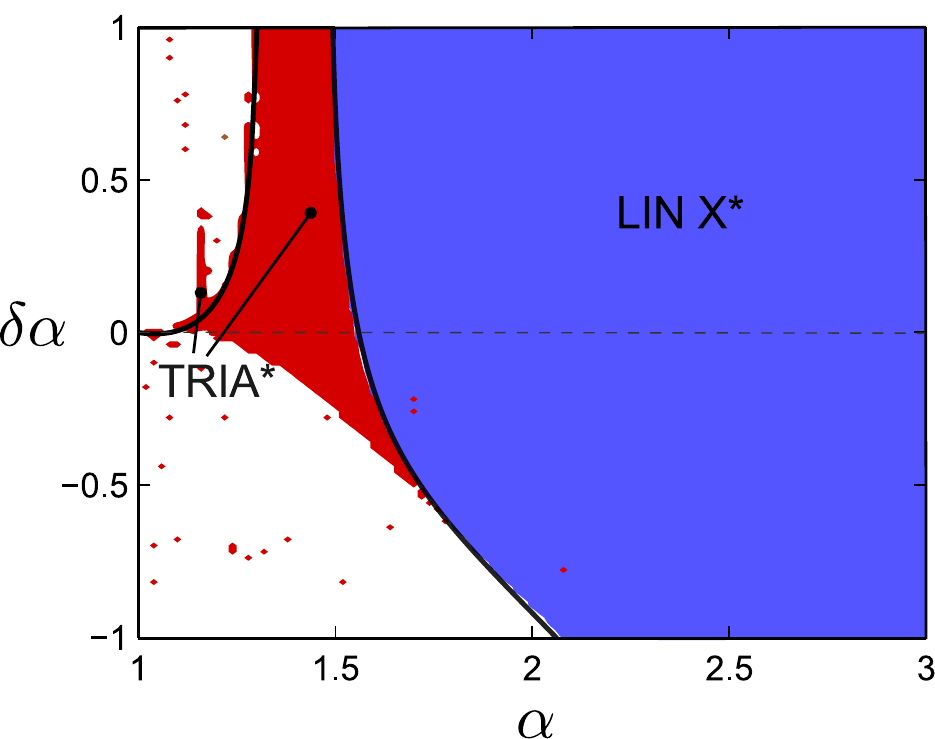}
\caption{
\label{f:outer_ion_excited}
(color online) Same as Fig. \ref{f:central_ion_excited} when one outer ion experiences the state-dependent potential. The labels ``LIN X*'' and ``TRIA*'' correspond respectively to a linear chain along the $x$-axis, and a triangular structure. The solid lines correspond to the stability boundaries in Eqs.~(\ref{eq:stability boundary 2})  and (\ref{eq:stability boundary 3}), the others were calculated numerically. The white regions are such that there is no stable structure with one outer ion excited.}
\end{figure}

The triangular structure TRIA* does not exhibit the symmetry properties of the zigzag structure. The corresponding equilibrium positions, as well as the stability boundary shown in Fig.~\ref{f:outer_ion_excited}, were calculated numerically using a Metropolis algorithm with random initial positions followed by a constraint minimization setting the asymmetry as a constraint. However, as the structures transform smoothly, one has to allow a certain tolerance on the constraint. On the other hand, the algorithm might run into a minimum corresponding to another structure satisfying accidentally the constraints. We believe this is the reason why scattered singular solutions are observed  in the stability diagram for the triangular configuration.

\end{document}